\documentclass[11pt,oneside]{amsart} \textwidth 4.7in \textheight 7.5in

\usepackage{citesort}
\usepackage{url}
\usepackage{amssymb}

\newcommand{\AFG}{Anderson-Fefferman-Graham}
\usepackage{cite}

\newcommand{\textcolor}[1]{}

\newcommand{\mcL}{{\mycal L}}

\newcommand{\bfg}{\bar g}
\newcommand{\bg}{\bar\g}
\newcommand{\bK}{\bar K}
\newcommand{\bhyp}{\,\,\overline{\!\!\hyp}}
\newcommand{\bmcM}{\,\,\,\,\overline{\!\!\!\!\mcM}}

\newcommand{\g}{\gamma}
\newcommand{\loc}{\mathrm{loc}}
\newcommand{\mcO}{{\mycal O}}
\newcommand{\mcU}{{\mycal U}}
\newcommand{\mcM}{{\mycal M}}
\newcommand{\hyp}{{\mycal S}}
\newcommand{\Scri}{{\mycal I}}
\newcommand{\scri}{{\Scri}}

\newcommand{\cH}{{\mathcal H}}

\newcommand{\cF}{{\mathcal F}}

\DeclareFontFamily{OT1}{rsfs}{}
\DeclareFontShape{OT1}{rsfs}{m}{n}{ <-7> rsfs5 <7-10> rsfs7 <10-> rsfs10}{}
\DeclareMathAlphabet{\mathscr}{OT1}{rsfs}{m}{n}

\newcommand{\eq}[1]{\eqref{#1}}

\newcommand{\tr}{\mathrm{tr}}

\newcommand{\bel}[1]{\begin{equation}\label{#1}}
\newcommand{\beaa}{\begin{eqnarray*}}
\newcommand{\bea}{\begin{eqnarray}}
\newcommand{\beal}[1]{\begin{eqnarray}\label{#1}}
\newcommand{\bean}{\begin{eqnarray}\nonumber}
\newcommand{\beadl}[1]{\begin{deqarr}\label{#1}}
\newcommand{\eeadl}[1]{\arrlabel{#1}\end{deqarr}}
\newcommand{\eeal}[1]{\label{#1}\end{eqnarray}}
\newcommand{\eead}[1]{\end{deqarr}}
\newcommand{\eea}{\end{eqnarray}}
\newcommand{\eeaa}{\end{eqnarray*}}

\newcommand{\be}{\begin{equation}}
\newcommand{\ee}{\end{equation}}

\DeclareFontFamily{OT1}{rsfs}{}
\DeclareFontShape{OT1}{rsfs}{m}{n}{ <-7> rsfs5 <7-10> rsfs7 <10->
rsfs10}{} \DeclareMathAlphabet{\mycal}{OT1}{rsfs}{m}{n}

\newcounter{mnotecount}[section]

\newcommand{\N}{{\Bbb N}}

\newcommand{\rmnote}[1]{}

\newcommand{\Ric}{\operatorname{Ric}}

\def\mysavedown#1{\edef\mysubs{\mysubs#1}}
\def\mysaveup#1{\edef\mysups{\mysups#1}}
\def\mydown#1{{\mytensor}_{\vphantom{\mysubs}#1}}
\def\myup#1{{\mytensor}^{\vphantom{\mysups}#1}}
\def\tensor#1#2{
  #1
  \def\mytensor{\vphantom{#1}}
  \def\mysubs{\relax}
  \def\mysups{\relax}
  \let\down=\mysavedown
  \let\up=\mysaveup
  #2
  \let\down=\mydown
  \let\up=\myup
  #2
  }

\newcommand{\R}{\mathbb R}

\renewcommand{\to}{\rightarrow}

\renewcommand{\phi}{\varphi}
\renewcommand{\epsilon}{\varepsilon}
\renewcommand{\hat}{\widehat}

\def\crn#1#2{{\vcenter{\vbox{
        \hbox{\kern#2pt \vrule width.#2pt height#1pt
           }
          \hrule height.#2pt}}}}

\renewcommand{\hbar}{{\overline h}}

\newcommand{\pre}[2]{{{\vphantom{#2}}^{#1}}\kern-.2ex{#2}}

\theoremstyle{plain}
\newtheorem{theorem}{\sc Theorem}[section]

\newtheorem{Remark}[theorem]{\sc Remark}
\newtheorem{Theorem}[theorem]{\sc Theorem}

\newtheorem{Proposition}[theorem]{\sc Proposition}

\theoremstyle{definition}

\theoremstyle{remark}

\numberwithin{equation}{section}

\date{\today}

\begin{document}

\title[Asymptotically simple even-dimensional space-times]
{Asymptotically simple solutions of the vacuum Einstein equations in even dimensions}

\author{Michael T. Anderson}
\address{Michael T. Anderson,  Dept. of Mathematics, S.U.N.Y. at Stony Brook, Stony Brook, N.Y. 11794-3651, USA
} \email{anderson@math.sunysb.edu}
\urladdr{http://www.math.sunysb.edu/$\sim$anderson}
\author{Piotr T. Chru\'sciel}
\address{Piotr T. Chru\'sciel\\
D\'epartement de math\'ematiques\\
Facult\'e des Sciences\\
Parc de Grandmont\\
F37200 Tours, France} \email{piotr@gargan.math.univ-tours.fr}
\urladdr{http://www.phys.univ-tours.fr/$\sim$piotr}
\thanks{Partially supported by NSF grant DMS 0305865  (MA), and the Erwin Schr\"odinger Institute,
the Vienna City Council, and the Polish Research Council grant KBN
2 P03B 073 24 (PTC)}

\begin{abstract}
We show that a set of conformally invariant equations derived from the
Fefferman-Graham tensor can be used to construct global solutions of vacuum
Einstein equations, in all even dimensions. This gives, in particular, a new,
simple proof of Friedrich's result on the future hyperboloidal stability of Minkowski
space-time, and extends its validity to even dimensions.
\end{abstract}

\maketitle

\section{Introduction}\label{section:intro}

  Consider the class of vacuum solutions to the Einstein equations
$(\mcM, g)$ in $n+1$ dimensions, which are future asymptotically simple,
i.e. conformally compact, in the sense of Penrose, to the future of a
complete Cauchy surface $(\hyp, \g)$. A natural method to try to
construct such space-times is to solve a Cauchy problem for the
compactified, unphysical space-time $(\mcM, \bar g)$, and then recover
the associated physical space-time via a conformal transformation.
However, a direct approach along these lines leads to severe
difficulties, since the conformally transformed vacuum Einstein equations
form, at best, a degenerate system of hyperbolic evolution equations,
for which it is very difficult to prove existence and uniqueness of solutions.

  Friedrich~\cite{Friedrich83,Friedrich:pune} has developed a method to overcome
this difficulty in $3+1$ dimensions, by introducing a system of ``conformal Einstein
equations'' whose solutions include the vacuum Einstein metrics and
which transforms naturally under conformal changes. A variation
upon Friedrich's approach, again in $3+1$ dimensions, has been
presented in~\cite{ChBNovello}.

  In this paper, we develop a different approach to this issue which,
besides its simplicity, has the advantage of working in all \emph{even}
dimensions. The method, carried out for vacuum space-times with
$\Lambda > 0$ in~\cite{AndersonCIE}, is based on use of the
Fefferman-Graham (ambient obstruction) tensor $\cH$, introduced in
~\cite{FG}. The tensor $\cH$ is a symmetric bilinear form, depending on a
metric $g$ and its derivatives up to order $n+1$, cf. Section~\ref{SsAFG} for
further discussion. It is conformally covariant, (of weight $n-1$)
and metrics conformal to Einstein metrics satisfy the system
\bel{AFG0}\cH = 0\;.\ee When $n = 3$, i.e. in space-time dimension
4, the Fefferman-Graham tensor is the well-known Bach tensor.

  The main result of the paper, Theorem~\ref{TAFG}, is the proof of the
well-posedness of the Cauchy problem for the equation \eq{AFG0},
for Lorentz metrics. This leads to a new proof of Friedrich's
result on the future ``hyperboloidal" stability of Minkowski
space-time~\cite{Friedrich} (see Theorem~\ref{Tstability}), and
extends the validity of this result to all even
dimensions.\footnote{Once most of the work on this paper was
completed we have been informed of the work by
Schimming~\cite{Schimming:Banach}, who has done a local analysis,
related to ours, of the Cauchy problem for the Bach equations in
dimension four. The application of his work to global issues, for
instance concerning the vacuum Einstein equations, seems not to have been
addressed. We are grateful to R.~Beig for pointing out that
reference to us.} As a corollary we additionally obtain existence
of a large class of non-trivial, vacuum, even dimensional
space-times which are \emph{asymptotically simple} in the sense of
Penrose, see Theorem~\ref{Tgex}.

 We further note that in~\cite{AndersonCIE} existence
of solutions of the Cauchy problem for \eq{AFG0} is obtained by
pseudo-differential techniques.  Here we show that the \AFG\ (AFG)
equations \eq{AFG0} can be solved using an auxiliary, first order,
symmetrisable hyperbolic system of equations. This shows that
\eq{AFG0} is a well posed evolutionary system, directly amenable
to numerical treatment. Thus, in space-time dimension four we
provide an alternative to Friedrich's conformal equations for the
numerical construction of global
space-times~\cite{HusaTuebingen,Friedrich:tuebingen,Frauendiener:tuebingen}.

  Our methods do not apply in odd space-time dimensions, where the
situation is rather different in any case, as one generically
expects polyhomogeneous expansions with half-integer powers of
$1/r$, where $r$ is, say, the luminosity distance,
compare~\cite{Lengard,ChLengardnwe,HollandsWald,HollandsIshibashi}.

\section{The  Anderson-Fefferman-Graham equations}
\label{SsAFG}

 Let, as before, $n+1$ denote space-time dimension, with
$n$ odd. The Fefferman-Graham tensor $\mathcal H$ is a conformally
covariant tensor, built out of the metric $g$ and its derivatives up
to  order $n+1$, of the form
\begin{equation} \label{e2.13} {\mathcal H}  =
(\nabla^{*}\nabla)^{\frac{n+1}{2}-2}[\nabla^{*}\nabla(P) +
\nabla^{2}(trP)] + \cF^n\;,
\end{equation} where
\begin{equation} \label{e2.14}
P =  \Ric_{g} - \frac{R_{g}}{2n}g\;,
\end{equation}
and where $\cF^n$ is a tensor built out of lower order derivatives
of the metric (see, e.g.,~\cite{GrahamHirachi}, where the notation
${\mathcal O}$ is used in place of ${\mathcal H}$). It turns out
that $\cF^n$ involves only derivatives of the metric up to order
$n-1$: this is an easy consequence of Equation~(2.4)
in~\cite{GrahamHirachi}, using the fact that odd-power
coefficients of the expansion of the metric $g_x$ in
\cite[Equation~(2.3)]{GrahamHirachi} vanish. (For $n=3,5$ this can
also be verified by inspection of the explicit formulae for
$\cF^3$ and $\cF^5$ given in~\cite{GrahamHirachi}.)

The system of equations
\bel{AFG} \cH = 0 \ee
will be called the \AFG\ (AFG) equations. It has the following
properties~\cite{GrahamHirachi}:
\begin{enumerate}
\item The system \eq{AFG} is conformally invariant: if $g$ is a solution, so
is $\varphi^2 g$, for any positive function $\varphi$.
\item If $g$ is conformal to an Einstein metric, then \eq{AFG} holds.
\item $\cH$ is trace-free.
\item $\cH$ is divergence-free.
\end{enumerate}
Recall that $\cH$ was originally discovered by Fefferman and
Graham~\cite{FG} as an obstruction to the existence of a formal power
series expansion for conformally compactifiable Einstein metrics, with conformal
boundary equipped with the conformal equivalence class $[g]$ of
$g$. This geometric interpretation is irrelevant from our point of
view, as here we are  interested in \eq{AFG} as an equation on its
own.

\section{Reduction to a symmetrisable hyperbolic system}\label{Sred}

Let $g$ be a Lorentzian metric, let $\nabla$
be a connection (not necessarily $g$-compatible), and let $\Box$
denote an operator
 with principal part $g^{\mu\nu}\nabla_\mu\nabla_\nu$ (acting perhaps on tensors).
 Let $u$ be a tensor field, and let $\nabla^{(k)} u$ denote any tensor formed from
the $k$-th order covariant derivatives of $u$. For $k\ge 0$ consider the system
of equations
 \bel{mainequ}\Box^{k+1} u = F(x,u,
\nabla u, \nabla^{(2)} u,\ldots, \nabla^{(2k+1)} u )\;,\ee for
some smooth $F$. Here we allow the coefficients of $\Box$ as well
as the connection coefficients to depend smoothly upon $x$ as well
as upon the collection of fields $(u, \nabla u, \nabla^{(2)}
u,\ldots, \nabla^{(2k+1)} u)$, in particular the metric $g$ is
allowed to depend (smoothly) upon those fields. We will assume
that \eq{mainequ} is invariant under diffeomorphisms, although
this is not necessary for some of our results below, such as local
existence and local uniqueness of solutions.

We want to show that solutions of \eq{mainequ} can
be found by solving a first order symmetric hyperbolic system of
PDEs. The idea of the proof can be illustrated by the following
example. Consider the equation
\bel{te1} \Box^2 u = 0\;.\ee Introducing
$$\psi^{(0)} = u\;,\qquad \psi^{(1)} = \Box u\;,$$
it is easily seen that solutions of \eq{te1} are in one-to-one
correspondence with solutions of the system
\bel{sh2}\Box\left(\begin{array}{c}
  \psi^{(0)} \\
  \psi^{(1)}\\
\end{array}\right)
=\left(\begin{array}{c}
  \psi^{(1)} \\
  0\\
\end{array}\right)
\;.\ee It is then standard to write a symmetrisable-hyperbolic
first order system so that solutions of \eq{sh2} are in one-to-one
correspondence with solutions of the first order system with
appropriate initial data (compare the calculations in the proof
below).

Some work is needed when we want to allow  lower order derivatives
  as in the right-hand-side of \eq{mainequ}:

\begin{Proposition}
\label{Preduc} There exists a symmetrisable hyperbolic first order
system
\bel{symhyp}P\Phi = H(\Phi)\;,\ee where $P$ is a linear first order operator and
$H$ does not involve derivatives of $\Phi$, such that every
solution of \eq{mainequ}, with $(\mcM,g)$ time orientable, satisfies \eq{symhyp}.
\end{Proposition}

\begin{proof} Let  $\{e_a\}_{a=0,\ldots,n}=\{e_0,e_i\}_{i=1,\ldots,n}$
be an orthonormal frame for $g$, with $e_0$ a globally defined unit
timelike vector; (such vector fields always exist on time
orientable manifolds). We set
\beal{varphikdef} &\varphi^{(j)}=\{\varphi^{(j)}_{a_1\ldots a_i}\}_{1\le i
\le 2(k-j)}\;, \ \mbox{where} \quad \varphi^{(j)}_{a_1\ldots a_i}
= e_{a_1}\cdots e_{a_i} \Box^ju\;,&\\\label{varphidef}
&\varphi=\{\varphi^{(j)}\}_{0\le j \le k}\;, &\\&
\psi=\{\psi^{(j)}\}_{0 \le j \le k}\;, \ \mbox{where} \qquad
\psi^{(j)}= \Box^j u\;. & \eeal{psidef} Let us derive a convenient
system of equations for $\varphi$. First,
\bel{eq1} \mbox{ for $1\le i\le 2(k-j)-1$} \qquad e_0\varphi^{(j)}_{a_1\ldots a_i}=\varphi^{(j)}_{0a_1\ldots
a_i}=L(\varphi^{(j)})\;,\ee where we use a generic symbol $L$ to
denote a linear map which may change from line to line. This gives
 evolution equations for those $\varphi^{(j)}_{a_1\ldots a_i}$'s
which have a  number of indices strictly smaller than the maximum
number allowed.

We continue by noting that, again for $1\le i+2j\le 2k-1$, we have
on the one hand
\beal{wvk} \Box\varphi^{(j)}_{a_1\ldots a_{i}}&=&-e_0\varphi^{(j)}_{0a_1\ldots a_{i}}+\sum_{\ell=1}^n
e_\ell\varphi^{(j)}_{\ell a_1\ldots a_{i}} +
L(\varphi^{(j)})\;,\eea and on the other
\beal{wvk2} \Box\varphi^{(j)}_{a_1\ldots a_{i}}&=&e_{a_1}\cdots e_{a_{i}}\Box \psi^{(j)} +
[\Box,e_{a_1}\cdots e_{a_{i}}]\psi^{(j)} \\\nonumber &=&
\varphi^{(j+1)}_{a_1\ldots a_{i}} + L(\varphi^{(j)})\;.\eea
Combining those two equations we obtain
\bel{eq23} e_0\varphi^{(j)}_{0a_1\ldots a_{i}}=\sum_{\ell=1}^n
e_\ell\varphi^{(j)}_{\ell a_1\ldots a_{i}} + L(\varphi^{(j)},
\varphi^{(j+1)}) \;.\ee Note that the condition $i+2j\le 2k-1$
implies $j<k$ so that \eq{eq23} can also be rewritten as
\bel{eq2n} e_0\varphi^{(j)}_{0a_1\ldots a_{i}}=\sum_{\ell=1}^n
e_\ell\varphi^{(j)}_{\ell a_1\ldots a_{i}} + L(\varphi) \;.\ee
Next, for $i+2j=2k-1$ and for $\ell$ running from $1$ to $n$ we
write
\bea \label{eq1n}  e_0\varphi^{(j)}_{\ell a_1\ldots a_i}&=&e_\ell \varphi^{(j)}_{0a_1\ldots
a_i}+[e_0,e_\ell]\varphi^{(j)}_{a_1\ldots a_{i}}
\\\nonumber &=&e_\ell\varphi^{(j)}_{0a_1\ldots
a_i}+L(\varphi^{(j)}) \;.\eeal{eq2} The rewriting of
\eq{eq2n}-\eq{eq1n} in the form
\bel{eq3}\left(
\begin{array}{cccc}
  e_0\varphi^{(j)}_{0a_1\ldots a_{i}} &
  -e_1\varphi^{(j)}_{1 a_1\ldots a_{i}} & \cdots & -e_n\varphi^{(j)}_{n a_1\ldots a_{i}} \\
  -e_1\varphi^{(j)}_{0 a_1\ldots a_{i}} & +e_0\varphi^{(j)}_{1 a_1\ldots a_{i}} & +0 & +0 \\
  \vdots & + 0 &  \ddots & \vdots \\
  -e_n\varphi^{(j)}_{0 a_1\ldots a_{i}} & +0 & \cdots & +e_0\varphi^{(j)}_{n a_1\ldots a_{i}} \\
\end{array}\right) = L(\varphi)
\ee makes explicit the symmetric character of \eq{eq2n}-\eq{eq1n}.
It is well known that this system is symmetrisable hyperbolic in
the sense of~\cite[Volume~III]{Taylor} when $e^0$ is a nowhere
vanishing vector field.\footnote{In fact, \eq{eq3} is symmetric
hyperbolic in a coordinate system with $e_0=\partial_t$ and
$e_it=0$. However, when $g$ depends upon $u$ and its derivatives
it is not useful to use such  coordinates, as the construction of
the Gauss coordinate system  leads to differentiability loss. In
any case Gauss coordinates are not well adapted to the proof of
existence of solutions when $g$ depends upon $u$.} This provides
the desired system of evolution equations for those
$\varphi^{(j)}_{a_0 a_1\ldots a_{i}}$'s which have the maximum
number of indices. (One could also use \eq{eq3} for any number of
indices, but \eq{eq1} is obviously simpler.)

If we write \eq{eq3} as
$$\mathring P \varphi = 0\;,$$
where $\mathring P$ is a linear first order operator, then the
derivatives $e_a \varphi$ satisfy a first order
symmetrisable-hyperbolic system of equations
\bel{eq3a}
\mathring P e_a \varphi =  L(\varphi, \nabla \varphi)\;, \qquad
L(\varphi, \nabla \varphi):=[\mathring P,e_a] \varphi\;. \ee

 The evolution
equations for $\psi$ are simply
\beal{eq4} \Box \psi^{(i)} &=& \psi^{(i+1)}\;,\qquad 0\le i \le
k-1\;, \\
\label{eq5} \Box \psi^{(k)} &=& F(x,\psi^{(0)},\varphi^{(0)},
\nabla \varphi^{(0)})\;,\eea where in \eq{eq5} we have expressed
the derivatives of $u$ appearing in \eq{mainequ} in terms of
$\varphi^{(0)}$ and $\nabla \varphi^{(0)}$ using \eq{varphikdef}.
By obvious modifications of the calculation starting at \eq{wvk}
and ending at \eq{eq3} one can rewrite the left-hand-side of
\eq{eq4}-\eq{eq5} as a first order symmetrisable hyperbolic
operator acting on the collection of fields $(\psi,\nabla
\psi):=\{(\psi^{(i)},\nabla \psi^{(i)})\}_{0\le i \le k}$. Setting
\bel{Phidef} \Phi= ( \psi, \nabla \psi, \phi, \nabla \varphi)\;,\ee
and letting $P$ be the linear part of the system of equations just
described, the proposition follows. \end{proof}

 The interest of Proposition~\ref{Preduc} relies in the fact, that it is standard to prove
 existence and uniqueness of solutions of \eq{symhyp} when the initial data
 for $\Phi$ are in $H^s$, $s\in \N$, for
 $s>n/2+1$, provided that $(\mcM,g)$ is globally hyperbolic.
 If $g$ does \emph{not} depend on $\nabla^{2k+1} u$, then the threshold can be
lowered\footnote{For $s>n/2+1$ the result follows
from~\cite[Volume~III, Theorem~2.3, p.~375]{Taylor}. However, when
the symmetric hyperbolic system has the structure considered here,
with $g$ \emph{not} depending upon $\nabla^{2k+1} u$,
the proof in~\cite{Taylor} applies for $s>n/2$.}
 to $s>n/2$.

  Now, not every solution of \eq{symhyp} will be a solution of
\eq{mainequ}. Let us show that
 appropriate initial data for \eq{symhyp} will provide the desired
 solutions.
When the space-time metric $g$ is independent of $u$,
 let $\hyp$ be a spacelike hypersurface in the space-time
 $(\mcM,g)$. We choose $e_0$ to be a unit time-like vector field normal to $\hyp$,
so that the $e_i$'s are tangential at $\hyp$, and we extend $e_0$
off $\hyp$ in some convenient way, which might vary according to
the context. Since \eq{mainequ} is an equation of order
$2k+2$, the associated Cauchy data consist of a set of tensor
fields $\{f_{(i)}\}_{i=0,\ldots,2k+1}$ defined on $\hyp$ which
provide initial data for $(e_0)^iu$ on $\hyp$:
\bel{ind}(e_0)^iu|_\hyp:=\underbrace{e_0\cdots e_0}_{\mbox{\scriptsize $i$
times}}u|_\hyp=f_{(i)}\;,\qquad 0\le i \le 2k+1\;.\ee For any
$i\ge0$ and $\ell\ge 0$ we can use \eq{ind} to calculate formally
$\psi^{(i)}|_{\hyp}$ and $\psi^{(i)}_{0}|_{\hyp}$  by replacing
each occurrence of $(e_0)^ju$ by  $f_{(j)}$, e.g.,
$\psi^{(0)}|_{\hyp}=f_{(0)}$, $e_0\psi^{(0)}|_{\hyp}=f_{(1)}$,
\beaa \psi^{(1)}|_{\hyp} &=&(\Box u)|_\hyp = \Big(-e_0e_0 u + \sum_{i=1}^n e_i
e_i u + \Gamma^\alpha e_\alpha u + \Gamma u\Big)\Big|_\hyp
\\ & = & -f_{(2)}+ \sum_{i=1}^n e_i
e_i f_{(0)} +  \Gamma^0 f_{(1)}+\sum_{i=1}^n\Gamma^i e_i f_{(0)}+
\Gamma f_{(0)}\;,\eeaa for some linear maps $\Gamma^\alpha$,
$\Gamma$ arising from the detailed structure of $\Box$, and so on.
We will write $g^{(i)}_{(0)}$ for the resulting functions
$\psi^{(i)}|_\hyp$ and $g^{(i)}_{(1)}$ for the resulting functions
$e_0\psi^{(i)}|_\hyp$, so that
$$\psi^{(i)}|_\hyp=g^{(i)}_{(0)}\;,\quad  e_0\psi^{(i)}|_\hyp=g^{(i)}_{(1)}\;.$$
Similarly we can calculate
$$\varphi^{(i)}_{0_1\ldots
0_\ell}|_{\hyp}:=\varphi^{(i)}_{\underbrace{{\mbox{\scriptsize
$0\ldots 0$}}}_{\mbox{\scriptsize $\ell$ factors}}}|_{\hyp}$$
where we replace each occurrence of $(e_0)^ju$ by  $f_{(j)}$, e.g.
$\varphi^{(0)}_{0_1\ldots 0_\ell}|_{\hyp}= f_{(\ell)}$. We will
write $h^{(i)}_{(\ell)}$ for the resulting functions, so that
$$\varphi^{(i)}|_{\hyp} =h^{(i)}_{(0)}=g^{(i)}_{(0)}\;, \quad
\varphi^{(i)}_{(1)}|_{\hyp} =h^{(i)}_{(1)}=g^{(i)}_{(1)}\;, \quad
\varphi^{(i)}_{0_1\ldots 0_\ell}|_{\hyp} =h^{(i)}_{(\ell)}\;.$$

When $g$ \emph{does} depend upon $u$, then the space-time will be built in the
process of solving the equations. In the simplest case of $g$ depending only upon $u$,
the procedure just described should be understood as follows:
the initial metric $g|_{\hyp}$ is determined by the initial data $f_0$. We choose an
orthonormal basis $\{e_i\}_{i=0,\ldots,n}$ for $g|_{\hyp}$, and interpret $e_0$ as the unit
normal to $\hyp$ in the space-time that will arise out of the initial data. Thus,
$f_{(1)}$ will be interpreted as the value of the normal derivative of $u$ at $\hyp$,
and so on, and the above considerations remain unchanged when this interpretation is used.

\begin{Proposition}
\label{Preduc2} Let $\Phi\in C(I,H^s(\mcO))$,  $s>n/2+2k+2$, $s\in
\N$, be a solution of \eq{symhyp} on a globally hyperbolic region
$I\times \mcO$ with initial data constructed as described above.
Then $u:=\psi^{(0)}$ is a solution of \eq{mainequ} and \eq{ind}.
\end{Proposition}

\begin{proof} From \eq{eq4} one
has $\psi^{(i)}=\Box^i \psi^{(0)}$ for $0\le i \le k$. It remains
to show that if $\varphi^{(0)}_{a_1\ldots a_i} = e_{a_1}\cdots
e_{a_i}\psi^{(0)}$, then \eq{eq5} will coincide with
\eq{mainequ}. This can be proved by a standard calculation. One sets
$$\chi^{(j)}_{a_1\ldots a_i} =\varphi^{(j)}_{a_1\ldots a_i} -
e_{a_1}\cdots e_{a_i}\psi^{(j)}\;,$$ and using \eq{symhyp} one
derives a system of equations which show that
$\chi^{(j)}_{a_1\ldots a_i}=0$ for the initial data under
consideration.

  However, the computations involved are avoided by
the following argument. Suppose, first, that $g$, $F$ and $\nabla$
are analytic functions of all their variables. Let us denote by
$f=\{f_{(j)}\}_{0\le j\le 2k+1}$ the initial data for
\eq{mainequ}; by an abuse of notation we will write $f\in H^s$ if
$f_{(j)}\in H^{s-j}$ for $0\le j\le 2k+1$. We note, first, that by
using an exhaustion of $I\times \mcO$ by compact subsets thereof
it suffices to prove the result when $I\times \mcO$ is a
conditionally compact subset of the domain of definition of the
solution. Let $f_n$ be any sequence of analytic initial data which
converges in $H^s(\mcO)$ to $f$. Let $u_n$ be the corresponding
solution of \eq{symhyp}; by stability all $u_n$'s will be defined
on $I\times \mcO$ for $n$ large enough. Similarly, the stability
estimates\footnote{Note that $I\times
\partial\mcO$ is non-timelike by global hyperbolicity, so that integration by parts
gives harmless contributions as far as energy estimates are
concerned.} for symmetric hyperbolic systems~\cite{KatoSH} prove
that $u_n$ is Cauchy in $C(I,H^{s}(\mcO))\cap C^1(I,H^{s-1}(\mcO))$.
The results in~\cite{AlinhacMetivier} show that $u_n$ is analytic
throughout $I\times \mcO$.

Let $\hat u_n$ be a solution of \eq{mainequ} on an open
neighborhood $\mcU_n$ of $\hyp$ in $I\times \mcO$ obtained by the
Cauchy-Kowalevska theorem. (Note that $\mcU_n$ could in principle
shrink as $n$ tends to infinity, but it is nevertheless open and
nonempty for each $n$.) Passing to a subset of $\mcU_n$ if
necessary we can without loss of generality assume that $\mcU_n$
is globally hyperbolic.

Now, uniqueness of the solutions of the Cauchy  problem for
\eq{symhyp} shows that $u_n$ coincides with $\hat u_n$ on
$\mcU_n$. Thus $u_n$ satisfies \eq{mainequ} there and thus, by
analyticity, everywhere. This shows that maximal globally
hyperbolic solutions of \eq{mainequ} with analytic initial data
are in one-to-one correspondence with maximal globally hyperbolic
solutions of \eq{symhyp} with the initial data constructed as
above. Then, for $H^s$ initial data, Proposition~\ref{Preduc2}
follows from continuity of solutions upon initial data for
\eq{symhyp}.

Finally, if the fields $g$, $\nabla$ and $F$ are smooth functions
of their arguments, they can be approximated by a sequence of
fields $g(n)$, $\nabla(n)$ and $F(n)$ which are analytic in their
arguments. The estimates for \eq{symhyp} just described can
similarly be used to show that solutions of the approximate
problem converge to solutions of the problem at hand both for
equation~\eq{symhyp} and \eq{mainequ}, which finishes the proof.
\end{proof}

{}From what has been said so far we obtain

\begin{Theorem}
\label{Tex} Let $ s>n/2+2k+2$, $s\in \N$. For any fields
$$f_{(i)}\in H^{s-i}_\loc(\hyp)\;,\qquad i=0,\ldots,2k+1\;,$$ there
exists a unique solution of \eq{mainequ} satisfying \eq{ind}. If
the metric $g$ does \emph{not} depend upon $\nabla^{2k+1} u$, then
$s>n/2+2k+1$ suffices.
\end{Theorem}

We note that in local coordinate systems $(t,x^i)$ on an open
neighborhood $\mcU$ of $\mcO\subset \hyp$ of the form
$\mcU=I\times \mcO$, with $\hyp\cap \mcU=\{t=0\}$ and
$\overline\mcO$-compact, the solutions are in
$$u \in \cap_{i=0}^{2k+1}C^{i}(I, H^{s-i}(\mcO))\;.$$
As usual, the lower
bound for the local time of existence of the solution does not
depend upon the differentiability class $s$, so in particular
smooth initial data provide smooth solutions.
 In addition the Cauchy problem for \eq{mainequ} is well-posed, in that
given a pair of initial data $f_{(i)}^{1}$, $f_{(i)}^{2}$ which
are close in $H_{\loc}^{s-i}(\hyp)$, then the solutions $u^{1}$,
$u^{2}$ are also close in $\cap_{i=0}^{2k+1}C^{i}(I,
H_{\loc}^{s-i}(\hyp))$.

When $g$ does not depend upon $u$, there exists a unique maximal
globally hyperbolic subset $\mcO$ of $\mcM$, with $\hyp$ being Cauchy for $\mcO$,
on which the solution exists. This is proved by the usual methods.

In the quasi-linear case one also has the existence of a maximal globally
hyperbolic development of the Cauchy data, giving a space-time $(\mcM,g)$.
This follows from the fact that the domains of dependence for the
system constructed above are determined by the light-cones of the
metric $g$, so that a proof along the lines
of~\cite{ChoquetBruhatGeroch69},
(compare~\cite{Chorbits,ChoquetYork79}), applies.

\section{The Cauchy problem for the AFG equations}\label{SCp}

   The Cauchy problem for \eq{AFG} has a similar form to that
for the Einstein equations. Since the system \eq{AFG} is of order $n+1$,
the initial data consist of an $n$-dimensional Riemannian manifold $(\hyp,\g)$,
$n=2k+1\ge 3$, with $n$ symmetric two-tensors $K^{(i)}$ prescribed
on $\hyp$. The tensor fields $K^{(i)}$ represent the $i$-th time derivative
of the metric $g$ in a Gauss coordinate system around $\hyp$. Thus, in a
neighborhood of $\hyp$, (or more precisely a neighborhood of a bounded
domain in $\hyp$), one may write
\bel{gauss}
g = -dt^{2} + \gamma(t), \ee
where $\gamma(t)$ is a curve of metrics on $\hyp$. Setting $e_{0} = -\nabla t$,
one has
\bel{Kdef} K^{(i)} =  \frac{1}{2}{\mathcal L}_{e_{0}}^{i}g|_{t = 0} =
\frac{1}{2}\partial_{t}^{i}\gamma(t)|_{t=0}. \ee
In particular $K = K^{(1)}$ is the extrinsic curvature
tensor of $\hyp$ in the final space-time $(\mcM, g)$.

  The set $(\g,K^{(1)},\ldots, K^{(n)})$ is not arbitrary, since the equations
\bel{ceq} \cH(e_{0}, \cdot) = 0, \ee
only involve $t$-derivatives of $g$ up to order $n$, and so induce $(n+1)$
equations on $(\g,K^{(1)},\ldots, K^{(n)})$. Because \eq{ceq} is
diffeomorphism invariant, this is most easily seen in the coordinates \eq{gauss}
where $g_{0\alpha} = -\delta_{0\alpha}$, so that \eq{ceq} only involves
$t$-derivatives of $g_{ab}$, $a, b \geq 1$, up to order $n$. The fact that
the constraint equations \eq{ceq} are preserved under the evolution follows
in the usual way from the equation $\delta \cH = 0$.

 To describe the system of $n+1$ constraint equations \eq{ceq} in more
detail\footnote{An explicit form of \eq{ceq} in space dimension $3$ can be found
in~\cite{Schimming:Banach}. The parameterisation of the initial data there is
rather different from ours.},
the Gauss-Codazzi equations for the embedding $\hyp \subset (\mcM,
g)$ are:
\bel{gc1} R_{\gamma} - |K|^{2} + H^{2} = R + 2\Ric(e_{0},e_{0}),  \ee
\bel{gc2} \delta K - dH = \Ric(e_{0}, \cdot), \ee
where $H = tr K$. In addition, in a Gauss coordinate system $(t,x^i)$
near $\hyp$, the Raychaudhuri equation gives
\begin{equation} \label{ray}
\partial_{t}H + |K|^{2} = -\Ric(e_{0}, e_{0}).
\end{equation}

  We first point out that the curvature scalar $R = R_{g}$ is determined
directly by the initial data; this is in constrast to the situation with the
Einstein equations, where $R$ is determined by the evolution
equations for the metric. Namely, the left side of \eq{gc1} is
determined by the initial data, as is $\Ric(e_{0}, e_{0})$, by
\eq{ray}. Thus $R$ is determined by $\gamma$ and $K^{(i)}$, for $i
= 1,2$.

  To describe the form of the ``scalar constraint equation''
$\cH_{\mu\nu}n^{\mu}n^{\nu} = 0$, $n = e_{0}$, note that from
\eq{e2.14} one has $\tr P = (n-1)R_g/2n$. Together with \eq{gc1}
and \eq{ray}, this leads to
\bean \Box P_{00}-\nabla_0\nabla_0 \tr P & = & \left(\frac
{1-n} n \right) \left( \Box + \nabla_0 \nabla_0\right)\partial_t H
\\\label{ll}
&& + \frac 1 {2n}\left( \Box + (1-n)
\nabla_0\nabla_0\right)R_\gamma + ...\;,\eea
where ``$...$" stands for terms which contain less derivatives of
\emph{the space-time metric}. The $t$-derivatives of the metric of order $4$
cancel out, as expected, and one easily finds that the equation
$\cH_{\mu\nu}n^\mu n^\nu=0$ takes the form
\bel{hoces}\Delta^{\frac{n-1}{2}} R_\gamma=\rho_n\;,\ee where, as before,
$R_\gamma$ is the curvature scalar of $\g$,  with $\Delta=D^k D_k$ the Laplace
operator of $\g$. Finally, $\rho_n$ is a functional of
$(\g,K^{(1)},\ldots, K^{(n)})$ which does \emph{not} involve
derivatives of \emph{initial data} of order $n+1$, while the
left-hand-side does. This shows in particular that \eq{ceq} is a
non-trivial restriction on the initial data.

It also follows from \eq{ll} that $\rho_n$ will contain terms of
the form \bel{ll2} \frac 1 n\Delta (\tr K^{(n-1)}) - D^k D^l
K^{(n-1)}_{kl} \;,\ee with other occurrences of $K^{(n-1)}$ there,
if any, being also linear with at most one space-derivative. Thus,
the scalar constraint equation $\cH_{\mu\nu}n^\mu n^\nu=0$ can be
viewed either as a non-linear equation of order $n+1$ which puts
restrictions on $\gamma$ in terms of the remaining data,  or as a
second order linear PDE  for the trace of
$K^{(n-1)}$.

One can similarly check that the equation $\cH_{0i}=0$ takes the
form of a linear first order PDE for $K^{(n)}$, with principal
part
\bel{ll3}D^i(K^{(n)}_{ij} -\frac 1n\tr K^{(n)}g_{ij}) \;,\ee
where  $D$ denotes the Levi-Civita connection of $\g$.

  A set $(\g,K^{(1)},\ldots, K^{(n)})$ will be called an \emph{initial data set}
if the fields $(\g,K^{(1)},\ldots, K^{(n)})$ satisfy the
constraint equations \eq{ceq}. The collection of initial data sets
is not empty, as every solution of the general relativistic
constraint equations solves \eq{ceq}.

\medskip

 The equations \eq{ceq} are preserved by the following family of
transformations, related to the conformal invariance of \eq{AFG}.
Suppose that the data set $(\g,K^{(1)},\ldots, K^{(n)})$ arises from a
space-time $(\mcM,g)$ satisfying \eq{AFG}. Then $(\g,K^{(1)},\ldots,
K^{(n)})$ satisfies \eq{ceq}, and if $\Omega$ is any strictly
positive function on $\mcM$, then the set $(\tilde \g,\tilde
K^{(1)},\ldots, \tilde K^{(n)})$ obtained on $\hyp$ from the
metric $\Omega^{2}g$ also satisfies \eq{ceq}.  For example, if we
set
\begin{equation} \label{omega}
\omega:=\Omega|_\hyp\;,\quad \omega^{(j)}:=
\underbrace{e_{0} \cdots e_{0}}_{\mbox{\scriptsize$j$ factors}}
(\Omega)|_\hyp\;,
\end{equation}
where $e_0$ is, e.g., a geodesic extension of $e_0$ off $\hyp$,
then it holds that
\bel{gKtr} \tilde \g = \omega^{2} \g\;,\qquad \tilde K^{(1)} =
\omega^{} K^{(1)} + \omega^{(1)}\g\;.\ee
Similar but more complicated transformation formulae hold for
$\tilde K^{(i)}$, $i \geq 2$, see Appendix~\ref{addon}. This leads
to a family of transformations preserving \eq{ceq} which are parameterized
by $n+1$ functions $\omega$, $\omega^{(j)}$, $j=1,\ldots, n$, on $\hyp$,
which are arbitrary except for the requirement that $\omega>0$.

Recall that under a conformal transformation of the space-time metric we
have
\bea \label{sccf} \qquad\qquad \tilde g_{ij} = \Omega^{\frac 4 {n-1}}g_{ij}\quad
\Longrightarrow \quad \tilde R=\Omega^{-\frac 4 {n-1}}\left(R
-\frac{4n}{(n-1)\Omega} \Box_g\Omega\right)
\;.\eeal{scalconftransf2} It follows from this formula that for
any given $\omega=\Omega|_\hyp>0$ and
$\omega^{(1)}=e_{0}(\Omega)|_\hyp$ one can choose $\omega^{(2)}$
so that $\Omega$ solves the linear wave equation
\bel{sccf3}\frac{4n}{(n-1)}\Box_g\Omega = R\Omega\ee when restricted to
$\hyp$. (This equation is globally solvable in globally hyperbolic
space-times, but this is irrelevant for the current discussion.
Note that solutions of \eq{sccf3} might sometimes develop zeros;
these are essential in the analysis of the vacuum Einstein
equations). Note as remarked above that the
curvature scalar $R$ is determined by the order 2 part of the
initial data set. Taking further $e_0$-derivatives shows that the
remaining $\omega^{(j)}$'s may be chosen so that the conformally
transformed curvature scalar $\tilde R$, together with its normal
derivatives up to order $n-2$, vanish at $\hyp$.

 Now the conformal and diffeomorphism invariance of \eq{AFG} requires a suitable
choice of gauge in order to obtain a well-posed system. As in~\cite{AndersonCIE}, we use
constant scalar curvature for the conformal gauge and harmonic coordinates for the
diffeomorphism gauge; the treatment of the conformal gauge is somewhat different
here than in~\cite{AndersonCIE}.

  Thus, we require first that
\bel{Rgc} R = 0\;.\ee
If \eq{Rgc} holds, then \eq{AFG} takes the form
\begin{equation} \label{e2.13a}
(\nabla^{*}\nabla)^{\frac{n-1}{2}}\Ric =- \cF^n\;,
\end{equation}
In harmonic coordinates $\{y^{\alpha}\} = \{(\tau, y^{i})\}$ with respect
to the conformal gauge \eq{Rgc}, one has
$$\Ric_{ab} = -\frac{1}{2}g^{\mu\nu}\partial_{\mu}\partial_{\nu} g_{ab} +
Q_{ab}(g,\partial g)\;,$$
where $Q$ is quadratic in $g$ and $\partial g$. Applying $(\nabla^{*}\nabla)^{(n-1)/2}$
to this and commuting $(\nabla^{*}\nabla)^{(n-1)/2}$ with the $\partial g$ terms in $Q$
shows that \eq{e2.13a} has the form
\begin{equation} \label{eharm}
\Box_{g}^{\frac{n+1}{2}}g_{\alpha\beta} = - \hat F_{\alpha\beta}^{n},
\end{equation}
where $\hat F_{\alpha\beta}^{n}$ still has the form
\eq{mainequ} and $\Box_{g} = g^{\mu\nu}\partial_{\mu}\partial_{\nu}$ acts on scalars.

  Finally, as initial data for the (gauge-dependent) variables $g_{0\alpha}$,
$0 \leq \alpha \leq n$, we choose
$$g_{0\alpha} = -\delta_{0\alpha}, \ \ {\rm on} \ \ \hyp.$$
The data $g_{ab}$ and $\partial_{\tau}g_{ab}$, $a, b \geq 1$, are determined by the
initial data $\gamma$, $K^{(1)}$. The derivatives $\partial_{\tau}g_{0\alpha}$
are then fixed by the requirement that the coordinates $\{y^{\beta}\}$ are
harmonic when restricted to $\hyp$, i.e.
\begin{equation} \label{harmcoord}
\Box y^{\beta} = \partial_{\alpha}g^{\alpha\beta} + \frac{1}{2}g^{\alpha\beta}g^{\mu\nu}
\partial_{\alpha}g_{\mu\nu} = 0 \ \ {\rm on} \ \ \hyp \;.
\end{equation}
The higher derivatives $\partial_{\tau}^{i}g_{\alpha\beta}$, $i \leq n$, on $\hyp$ are then
determined by the given initial data $(\g,K^{(1)},\ldots, K^{(n)})$ and by setting to zero
the $\tau$-derivatives of \eq{harmcoord} up to order $n-1$.

 These choices lead to the following:

\begin{Theorem}
\label{TAFG}  Consider any class $(\hyp,[\g,K^{(1)},\ldots, K^{(n)}])$
satisfying the constraint equations \eq{ceq}, with
$$(\g,K^{(1)},\ldots, K^{(n)})\in H^s_\loc(\hyp)\times
H^{s-1}_\loc(\hyp)\times \cdots \times H^{s-n}_\loc(\hyp)\;,$$ $
s>n/2+n+1$,  $s\in \N$, where $(\hyp,\g)$ is a Riemannian metric
and where the $K^{(i)}$'s are symmetric two-covariant tensors; the
equivalence class is taken with respect to the transformations of
the data discussed above.

  Then there exists a unique maximal
globally-hyperbolic conformal space-time $(\mcM, [g])$ satisfying
\eq{AFG}, where $[\cdot]$ denotes the conformal class, and an
embedding $$i:\hyp \to \mcM\;,$$ for which $\g$ is the metric
induced on $\hyp$ by $g$, with $K^{(i)}$ given by \eq{Kdef}. One
can always choose local representatives of $[g]$ by imposing
\eq{Rgc}. Moreover, the Cauchy problem with such initial data is
well-posed in $H^{s}_\loc(\hyp)\times H^{s-1}_\loc(\hyp)\times
\cdots \times H^{s-n}_\loc(\hyp)$.
\end{Theorem}

\begin{Remark}{\rm The inequality
$s>n/2+n$ suffices for existence of unique solutions in local
coordinate patches in Theorem~\ref{TAFG}. One expects that the use
of  local foliations with prescribed mean curvature and space-harmonic
coordinates as in \cite{AnderssonMoncriefAIHP} should allow one to
lower the threshold $s>n/2+n+1$ to $s>n/2+n$ in this result.}
\end{Remark}
\begin{Remark}{\rm The conformal space-time $(\mcM,[g])$ is smooth if
the initial data are. Similarly, real-analytic initial data lead to
real-analytic solutions.}
\end{Remark}

\begin{proof} Given any initial data set
$[(\g,K^{(1)},\ldots, K^{(n)})]$ satisfying the constraint equations, by
using the functions $\omega^{(j)}$, $j=2,\ldots,n$ from \eq{omega}, one
can adjust the tensor fields $K^{(i)}$, $i=2,\ldots,{n}$, so that
$R_{g}$, together with its transverse derivatives up to order
$n-2$, vanish on $\hyp$, (see the discussion following \eq{sccf3}).
Note that this holds for any $\g \in [\g]$, so that $\omega > 0$ and
$\omega^{(1)}$, while fixed, are otherwise freely specifiable.
Solving \eq{eharm} with this initial data, as described in
Section~\ref{Sred}, one obtains a collection of space-time
coordinate patches with a solution of \eq{eharm} there when
$s>n/2+n$. We recall again that the Cauchy problem for \eq{eharm}
is well-posed in the $H_{\loc}^{s}$ spaces above.

The argument that these gauge choices are preserved, so that \eq{Rgc}
holds and the coordinates $y^{\alpha}$ remain harmonic in
these local space-time coordinate patches generated by \eq{eharm},
is rather similar to the one for the Einstein
equations~\cite{ChBActa}; we give details because of some
differences in the analytical tools used.

First recall that \eq{eharm}
can be written in the form
\begin{equation} \label{eharm1}
\Box_{g}^{\frac{n-1}{2}}(R_{\mu\nu}-\nabla_\mu \lambda_\nu
-\nabla_\nu \lambda_\mu) = -  \cF^{n}_{\mu\nu}\;,
\end{equation}
where
$$\lambda^\mu:= -\frac{1}{2}\Box_g y^\mu\;.$$
As $g$ solves \eq{eharm1}, its obstruction tensor equals
\begin{equation} \label{cH}
\cH_{\mu\nu}=
(-1)^{\frac{n-1}2}\left\{\Box_{g}^{\frac{n-1}{2}}(\nabla_\mu
\lambda_\nu + \nabla_\nu \lambda_\mu - \frac1{2 n} R_g g_{\mu\nu}
)-\frac{n-1}{2n}\Box_{g}^{\frac{n-3}{2}} \nabla_\mu \nabla_\nu
R_g\right\}\;.
\end{equation}
The divergence identity $\nabla_\mu \cH^{\mu\nu}=0$ gives then an
equation involving $\lambda$ and $R_g$:
\bel{feq1} \Box_{g}^{\frac{n-1}{2}}\left\{\Box \lambda_\nu +
\nabla_\nu (\nabla^\mu\lambda_\mu-\frac 12 R_g)\right\}= 
\mbox{lower order commutator terms}\;.\ee Since
$\cH$ is trace-free, we further have from \eq{cH}
\bel{feq2} \; \Box_{g}^{\frac{n-1}{2}}(\nabla^\mu \lambda_\mu - \frac{1}{2} R_g) =0 \;. \ee
Because $R_g$ vanishes to order $n-2$ at $\hyp$, and $\lambda_{j}$
vanishes to order $n-1$ at $\hyp$ by the discussion following
\eq{harmcoord}, the initial data for this equation vanish.
It follows, for instance from the work in Section~\ref{Sred}, that
\bel{feq5} R_g = 2\nabla^\mu\lambda_\mu\;.\ee
 This can be
used to rewrite \eq{feq1} as
\bel{feq3}
\Box_{g}^{\frac{n+1}{2}}\lambda_\nu = \mbox{lower order commutator
terms}\;.\ee In this last equation all commutator terms in
\eq{feq1} that involve gradients of $R_g$ have been replaced by
derivatives of $\lambda$ using \eq{feq5}.

Since $g$ satisfies the constraint equation \eq{ceq} at $\hyp$,
\eq{feq5} and \eq{cH} imply that $\lambda_{j}$ vanishes to order
$n$ at $\hyp$, so that \eq{feq3} has vanishing initial data. This
system has the form considered in Section~\ref{Sred}, so we
conclude that $\lambda$ vanishes. Hence, the coordinates
$y^{\alpha}$ remain harmonic, $R_g=0$ by \eq{feq5}, and so $\cH =
0$ as well, as desired.

The usual procedure, as used in the context of the Cauchy problem
for Einstein's equations, then allows one to patch the solutions
together provided $s>n/2+n+1$. An argument as in~\cite{ChoquetBruhatGeroch69}
leads to a unique, (up to diffeomorphism) maximal globally hyperbolic manifold
$(\mcM,[g])$, with $[g]$ satisfying \eq{e2.13a}, with an embedding
$i:\hyp\to\mcM$, with the desired initial data on $i(\hyp)$.

The well-posedness statement follows immediately from the
discussion following Theorem~\ref{Tex}.
\end{proof}

\begin{Remark} \label{global}
{\rm In general, there will not be a global smooth gauge for $(\mcM, [g])$
in which $R_{g} = 0$. The local coordinate patches where $R_{g} = 0$ need not
patch together smoothly, preserving $R_{g} = 0$, since the initial data
$\omega > 0$, $\omega^{(1)}$ on local space-like slices $\hyp$ are freely
chosen, and so not uniquely determined.

To see this in more detail, consider for example the de Sitter space-time
$\mcM = {\mathbb R}\times S^n$, with metric
$$g_{dS} = -dt^2 + \cosh^{2}t\;g_{S^{n}(1)}.$$
This is a geodesically complete solution of the Einstein equations with
$R_{g} = n(n+1)$, and so satisfies \eq{AFG}. The linear wave equation
\eq{sccf3} with initial data $\Omega = c_1$, $\partial_{t}\Omega = c_2$
on $\hyp = \{t = 0\}$ has a global solution on $\mcM$ given by
$$\Omega = -\frac{n-1}{4}\sinh t + d_{2}\int_{0}\frac{1}{\cosh^{n}t}dt + d_{1},$$
for suitable $d_{1}, d_{2}$. One sees that there are no values of $c_1, c_2$
for which $\Omega > 0$ everywhere on $\mcM$, so that there is
no natural global $R = 0$ gauge for $(\mcM, [g_{dS}])$ with such initial data.

Consider for example the solution $\Omega = -\frac{n-1}{4}\sinh t$, giving an
$R = 0$ gauge in the region $t < 0$, which does not extend to $\{t = 0\}$. For
$t = -\epsilon$ small, the induced metric on $\hyp_{-\epsilon} = \{t = -\epsilon\}$
is the round metric $\g_{\delta}$ on $S^{n}$ of small radius $\delta =
\delta(\varepsilon)$. To obtain an $R = 0$ gauge starting at $\hyp_{-\epsilon}$
which extends up to and beyond $\{t = 0\}$, one must choose $\omega$ to be a
large constant. This causes a discontinity in the choice of gauge for the metric,
but not in the structure of the conformal class. }
\end{Remark}

\begin{Remark} \label{Rc}
{\rm One may also construct the maximal solution $(\mcM, [g])$ by means of local
gauges satisfying
\begin{equation} \label{c0}
R_{g} = c_0
\end{equation}
in place of the scalar-flat gauge \eq{Rgc}, for any $c_0 \in {\mathbb R}$.
The proof of this is the same as before, noting from the form of \eq{sccf}
that given $\omega$ and $\omega^{(1)}$ on $\hyp$, one can find $\omega^{(2)}$ on
$\hyp$ such that $\tilde R = c_0$ on $\hyp$.

For example, the de Sitter space-time is a geodesically complete solution in
the gauge $R_g = n(n+1)$. However, it is well-known that $(\mcM, g_{dS})$
conformally compactifies to the bounded domain in the Einstein static cylinder
$g_{E} = -dT^2 + g_{S^{n}(1)}$ where $T \in (-\frac{\pi}{2}, \frac{\pi}{2})$.
The metric $g_{E}$ is of course also a solution of \eq{AFG} with $R_{g_E}  = n(n-1)$,
which is thus a globally hyperbolic extension of $(\mcM, g_{dS})$, since
$T \in (-\infty, \infty)$; it is easily seen that this is the maximal solution.

Thus, the choice $R_{g_{E}}  = n(n-1)$ gives a global gauge for $[g_{dS}]$. The
de Sitter metric itself, with gauge $R_g  = n(n+1)$, has conformal factor
$\Omega$ relative to $g_{E}$ blowing up to $\infty$ as $T \rightarrow \pm \frac{\pi}{2}$.
To obtain an extension of $[g_{dS}]$ past the range $(-\frac{\pi}{2}, \frac{\pi}{2})$ of
$g_{dS}$ requires a rescaling of the large $\Omega$ factor to a factor of unit size. }
\end{Remark}

\section{AFG equations vs. Einstein equations}\label{SAFGvsE}

Consider an initial data set $(\hyp,\g,K)$ for the vacuum Einstein
equations in $n+1$ dimensions, $n$ odd. Thus $(\hyp,\g)$ is a Riemannian manifold,
and the pair $(\g,K)$ satisfies the vacuum constraint equations with
cosmological constant $\Lambda \in \R$. Using Einstein's equations
one can formally calculate the derivatives
\begin{equation} \label{KEdef}
K^{(i)}_{kl}:= \frac{1}{2}\partial_t^{i}\gamma_{kl}|_{t=0}\;,\qquad 1\le i \le n
\end{equation}
in a hypothetical Gauss coordinate system near $\hyp$ in which the
space-time metric $g$ takes the form $-dt^{2} + \gamma(t)$, as in
\eq{gauss}. This gives

\begin{Proposition}
\label{PEv} The initial data set for \eq{AFG} so obtained solves
the constraints \eq{ceq}, and any such globally hyperbolic solution of
\eq{AFG} given by Theorem~\ref{TAFG} is conformally Einstein.
\end{Proposition}

\begin{proof} Let $g$ be the associated maximal globally hyperbolic
solution of the vacuum Einstein equations. Then $g$ also solves \eq{AFG},
and the result follows from the uniqueness part of Theorem~\ref{TAFG}.
\end{proof}

 A space-time with boundary $(\bmcM,\bfg)$
is said to be a \emph{conformal completion at infinity} of a
space-time $(\mcM,g)$ if the usual definition of Penrose is
satisfied; that is, there exists a diffeomorphism $\Phi$ from
$\mcM$ to the interior of $\bmcM$ and a function $\Omega:\bmcM\to
\R^+ \cup \{0\}$, which is a defining function for the boundary
$$\Scri:=\partial \bmcM\;,$$ such that
$$ g=\Phi^*(\Omega^{-2}\bfg)\;.$$
A conformal completion $(\bmcM,\bfg)$ is $H^{s}$ smooth if $\bfg \in
H^{s}(\bmcM)$, (in suitable local coordinates for $\bmcM$).

 A set $(\bhyp,\bg,\bK,\omega,\omega^{(1)},\ldots,\omega^{(n)})$ is said to
 be a \emph{smooth conformal
 completion at infinity} of a general relativistic initial data set $(\hyp,\g,K)$
if  $(\bhyp,\bg)$ is a Riemannian manifold with boundary, with
$\bhyp=\hyp\cup\partial\hyp$,  and with
 $$\omega,\omega^{(j)}:\bhyp\to\R$$ being smooth-up-to-boundary functions
 such that $\omega$ is a defining function
 for the boundary $$\dot\hyp:=\partial\bhyp\;,$$ with
 \bel{cv1}\bg = \omega^{2} \g\ee
on $\hyp$. Finally $\bK$ is a smooth-up-to-boundary symmetric
tensor field on $\bhyp$ such that the equation
\bel{cv2}
\bar K = \omega^{} K + \frac{\omega^{(1)}}{\omega^2} \bar \g\ee
holds on $\hyp$ (compare \eq{gKtr}). We further assume that the
functions $\omega^{(i)}$, $2\le i\le n$, are such that the
fields $\bar K^{(i)}$, calculated using the $K^{(i)}$'s as in
\eq{KEdef}, and the functions $\omega^{(i)}$, defined before \eq{gKtr},
can be extended by continuity to smooth tensor fields on $\bhyp$.

A conformal completion will be said to be $H^s$ if $\bg\in H^{s}(\bhyp)$
and $\bK^{(i)}\in H^{s-i}(\bhyp)$, $i=1,\ldots,n$.

 The above conditions are clearly  necessary for the existence of a smooth
conformal completion \emph{\`a la Penrose} of the maximal globally
hyperbolic development of $(\hyp,\g,K)$; we will see shortly that
they are also sufficient. We emphasise, however, that
Equations~\eq{cv1} and \eq{cv2} alone, together with the
requirement of smoothness of $\bg$ and $\bK$ are \emph{not}
sufficient for the existence of such space-time completions.
Indeed, it follows from the results in~\cite{AC} that, in
dimension $3+1$, the requirement of smoothness up-to-boundary of
the $\bK^{(i)}$'s, $i\ge 2$, imposes further constraints on $\bg$
and $\bK$. It would be of interest to work out the explicit form
of those last conditions, analogously to \cite{AC}, in all
dimensions.

We have the following, conformal version of Proposition~\ref{PEv};
it allows one to repeat several constructions of Friedrich
(see~\cite{Friedrich:tuebingen} and references therein) for vacuum
space-times with vanishing cosmological constant in all even
space-time dimensions:

\begin{Theorem}
\label{TcE} Let $(\hyp,\g,K)$ be a general relativistic vacuum initial
data set, $\Lambda = 0$, which admits an $H^s$ \emph{conformal completion at
infinity}, $ s>n/2+n+1$, with $s\in \N$ and $n$ odd. Then there exists an
$H^{s}$ space-time with boundary $(\bmcM,\bfg)$, equal to the conformal
completion at infinity of the unique maximal development $(\mcM,g)$ of
$(\hyp,\g,K)$, so that $g=\Omega^{-2}\bar g$ on  $\mcM$, and
$$\Scri\supset \dot \hyp\;.$$
\end{Theorem}

\begin{proof}
The proof is essentially identical to that of Proposition~\ref{PEv}. Consider
the initial data $(\g,K^{(1)},\ldots, K^{(n)})$ as constructed at the beginning
of this section. One can conformally transform them to initial data
$$\bK^{(i)}_{kl}:=\frac{1}{2}\partial_s^{i}\bg_{kl}|_{t=0}\;,\qquad 1\le i \le n$$
in a hypothetical Gauss coordinate system near $\bhyp$ in which the
space-time metric $\bfg$ takes the form $\bfg=-ds^2+\bg(s)$. Here we use a
normalisation as in the proof of Theorem~\ref{TAFG} or Remark~\ref{Rc},
requiring the vanishing (or constancy) of $R_{\bar g}$. Theorem~\ref{TAFG}
provides a solution of this Cauchy problem, while the fact that this solution
is conformally Einstein follows from Proposition~\ref{PEv}.

The choice of the conformal factor $\Omega$ transforming $({\bmcM}, [\bar g])$
to the vacuum Einstein solution $(\mcM, g)$ of course depends on the choice of
gauge for $[\bar g]$; we describe here how $\Omega$ is determined at least in the
natural settings corresponding to \eq{c0}. Suppose that $\bar \gamma = \omega^{2}\gamma$
is a geodesic compactification of $(\hyp, \g)$, so that for $x$ near $\dot \hyp$,
$\omega (x) = dist_{\bar \g}(\dot \hyp, x)$. Such a compactification is uniquely
determined by the choice of a boundary metric on the boundary $\dot \hyp$. Now the
value of $\omega^{(1)}$, at the zero level set of $\omega$ is determined by
the initial data, (compare Eq. (3.13) of ~\cite{AC} in dimension $3+1$; an obvious
modification of that equation holds in all dimensions). Changing time-orientation
if necessary, one will have $\omega^{(1)} = -1$ at $\dot \hyp$ and we extend
$\omega^{(1)}$ to a neighborhood of $\dot \hyp$ in $\hyp$ to have the same value.
These data determine the compactification of the initial data set $(\hyp, \g, K)$.

Hence, the proof of Theorem~\ref{TAFG} gives a unique local solution $\bar g$
of \eq{AFG} in the conformal class $[\bar g]$, with $R_{\bar g} = c_{0}$, for
any given $c_0$, satisfying the initial conditions. Let $\phi = \Omega^{(n-1)/2}$,
and set $g = \Omega^{-2}\bar g = \phi^{-4/(n-1)} \bar g$. Then $\phi$, (and so $\Omega$),
is uniquely determined in a chart for $\tilde{\mcM}$ containing a portion of $\dot \Scri$
by the requirement that $\phi$ solves the linear wave equation
$\frac{4n}{n-1}\Box_{\bar g}\phi - R_{\bar g}\phi = 0$, with initial data
$\phi = \omega$, $\partial_{s}\phi = \omega^{(1)} = -1$ on $\hyp$ near $\dot \hyp$,
(where $s$ is the Gaussian coordinate).

  Given such a solution $\Omega$, let $\Scri$ be the connected component of the set
$\{\Omega = 0\;,\ d\Omega \ne 0\}$ intersecting $\dot \hyp$. Since, by construction,
$Ric_{g} = 0$, and $\bar g$ is smooth up to $\Scri$, standard formulas for the Ricci
curvature under conformal changes show that $\Omega$ has the usual structure on $\Scri$,
in that $\nabla \Omega$ is null, and $\nabla^{2}\Omega$ is pure trace, on $\Scri$.
\end{proof}

\begin{Remark} \label{dScase}
{\rm A version of Theorem 5.2, and its proof, also holds for de Sitter-type vacuum
solutions of the Einstein equations, where $\Lambda > 0$. In this case, the completion
is at future or past space-like infinity $\Scri^{+}$ or $\Scri^{-}$; the Cauchy data
for $({\bmcM}, \bg)$ at $\Scri^{+}$, consist of the two undetermined terms
$g_{(0)}$, $g_{(n)}$ in the formal Fefferman-Graham expansion for vacuum Einstein
solutions with $\Lambda > 0$. This gives an alternate proof of one of the results of
~\cite{AndersonCIE}.

The case $\Lambda < 0$ leads to initial-boundary value problems. While it is clear
that a generalization of Friedrich's analysis of this case ~\cite{Friedrich:aDS}
should exist, precise statements require further investigation. }

\end{Remark}

\section{Applications to semi-global and global stability of  general relativistic
initial value problems in all even space-time dimensions}

 Let $(\hyp, \g_{0})$ be the Poincar\'e metric on the $(n+1)$-dimensional ball,
with $n \geq 3$ and $n$ odd. Setting
$$K_{0} = \g_{0},$$
the set $(\hyp, \g_{0}, K_{0})$ is an initial data set for the vacuum Einstein
equations, denoted \emph{standard hyperboloidal initial data}. The maximal
globally hyperbolic development $(\mcM, g_{0})$ of $(\hyp, \g_{0}, K_{0})$ is
given by
\begin{equation} \label{cone}
g_{0} = -d\tau^{2} + \tau^{2}\g_{0},
\end{equation}
for $\tau \in {\mathbb R}^{+}$, with $\hyp = \{\tau = 1\}$.
This space-time is the interior of the future light cone about a
point in Minkowski space-time (the ``Milne universe"). With
respect to the standard smooth conformal compactification of
Minkowski space-time as a bounded domain in the static Einstein
cylinder,
\begin{equation} \label{Estatic}
\bar g_{0} = -dT^{2} + dR^{2} + \sin^{2}R  \ g_{S^{n-1}(1)},
\end{equation}
one has $\bK^{(i)} = 0$, $1 \leq i \leq n$; the hypersurfaces
$\{\tau = const\}$ correspond to the level-sets $\{T = const\}$,
and so are totally geodesic. As noted in Remark~\ref{Rc}, this
choice of gauge is global and satisfies $R_{g_{0}} = n(n-1)$.

  As an example of application of the results of the previous section, one now
easily obtains:

\begin{Theorem}
\label{Tstability} Let $\hyp$ be an $n$-dimensional open ball, $n$
odd, and consider a general relativistic initial data set $(\hyp,
\g, K)$ which admits an $H^{s}$ \emph{conformal completion at
infinity}, $s>n/2+n+1$, $s\in \N$. Then there exists $\epsilon =
\epsilon(n)
> 0$ such that if the associated data
$(\hyp,[(\bg,\bK^{(1)},\ldots, \bK^{(n)})])$ are $\epsilon$-close
in $H^s\times\ldots\times H^{s-n}$ to the data $(\hyp,
[(\bar\gamma_{0}, 0, \ldots, 0)])$ associated to standard
hyperboloidal initial data, then the maximal globally hyperbolic
development $(\mcM, g)$ of $(\hyp,\g,K)$ is causally geodesically
complete to the future.

  The $H^{s}$ conformal compactification $(\bmcM, \bar g)$ is $H^{s}$ close to
$(\bmcM, \bar g_{0})$, and extends to a larger $H^{s}$ space-time
containing a regular future time-like infinity $\iota^{+}$ for
$(\bmcM, \bar g)$.
\end{Theorem}
\begin{proof}
The standard space-time $(\mcM, g_{0})$ has a conformal compactification to
a bounded domain ${\mathcal D}$ in the static Einstein cylinder \eq{Estatic},
where ${\mathcal D}$ corresponds to the range of parameters $T+R \in [0,\pi]$,
$T-R \in [0,\pi]$, $R \geq 0$. Future null infinity $\Scri^{+}$ is given by
$\Scri^{+} = \{T+R = \pi\}, T \in (\frac{\pi}{2}, \pi)$, with future time-like
infinity $\iota^{+} = \{T = \pi, R = 0\}$. The future development of
$(\hyp, \g_0, K_0)$ corresponds to the domain ${\mathcal D}^{+} = {\mathcal D}
\cap \{T \geq \frac{\pi}{2}\}$.

  Clearly, the compactification $({\mathcal D}, \bar g_{0})$ extends smoothly
to a neighborhood $\tilde{\mathcal D}$ of ${\mathcal D}$ as a globally
hyperbolic solution of \eq{AFG}. The Cauchy data for such an extension are an
extension of the standard Cauchy data $(\hyp,[(\bar\gamma_{0}, 0, \ldots, 0)])$
past the boundary $\dot \hyp$. Similarly, the initial data
$(\hyp,[(\bg,\bK^{(1)},\ldots, \bK^{(n)})])$ for $(\mcM, \bar g)$ extend in $H^{s}$
past a neighbhorhood of the boundary $\dot \hyp$ and generate a maximal globally
hyperbolic space-time $(\tilde{\mcM}, \tilde g)$, satisfying \eq{AFG}. By the
Cauchy stability associated with Theorem~\ref{TcE}, for $\epsilon$ small, the
solution $(\tilde{\mcM}, \tilde g)$ is close in $H^{s}$ to $(\tilde{\mathcal D},
\bar g_{0})$, and in particular is an $H^{s}$ extension of $(\mcM, \bar g)$, where
$\mcM = \{\Omega > 0\}$ in $\tilde{\mcM}$. This shows that, to the future of $\hyp$,
$(\mcM, g)$ has an $H^{s}$ conformal completion, which extends in $H^{s}$ to a
neighborhood of $\Scri$ and $\iota^{+}$. This gives the result.
\end{proof}

Note that in dimension $3+1$ the mere requirement
$$[(\bg,\bK^{(1)},\ldots, \bK^{(n)})]\in H^s\times\ldots\times H^{s-n}\;,\quad s>n/2+n+1\;,$$
regardless of any smallness condition, forbids solutions which have logarithmic terms  with
small powers of $1/r$ in polyhomogeneous expansions. Thus, (similarly to the results of
Friedrich), the above theorem applies for non-generic initial data sets only.

Using Corvino-Schoen type constructions together with the above
stability result, as in~\cite{ChDelay2}, one obtains:

\begin{Theorem}
\label{Tgex} There exists an infinite dimensional space of
complete, asymptotically simple globally hyperbolic solutions of
the Einstein vacuum equations in all even dimensions $n+1 =
2(k+1)$, $n \geq 3$. Thus, such solutions are geodesically
complete both to the future and past, and have a smooth conformal
completion at infinity.
\end{Theorem}

\begin{proof} The Corvino-Schoen gluing
technique~\cite{CorvinoSchoenprep,Corvino} can be used, as
in~\cite{ChDelay2,ChDelay}, to construct static, parity-symmetric
initial data on $\R^n$, for any $n\ge 3$, which are Schwarzschild
with $m\ne 0$ outside a compact set, and which are as close to the
Minkowskian data as desired. The resulting maximal globally
hyperbolic space-time then contains smooth hyperboloids, close to
standard hyperbolic initial data, as in Theorem~\ref{Tstability},
both in the future and in the past. In even space-time dimensions
the result follows as in the proof of Theorem~\ref{Tstability}.
\end{proof}

We note that all the space-times constructed in the proof of
Theorem~\ref{Tgex} possess a ``complete  $\scri$"; this should be
understood as completeness of generators of $\scri$ in the
zero-shear gauge, compare~\cite{GerochHorowitz}.\footnote{We are
grateful to H.~Friedrich for useful discussions concerning this
point.}

 One expects the above
construction to generalise to initial data which are stationary,
asymptotically flat outside of a spatially compact set (rather
than exactly Schwarzschild there). This would require proving that
the resulting  space-times have smooth conformal compactifications
near $\iota^0$, (in space-time dimension four this follows
from~\cite{SimonBeig,Dain:2001kntwice,Damour:schmidt}), and
working out ``reference families of metrics" needed for the
arguments in~\cite{ChDelay}. Those results are very likely to
hold, but need detailed checking; note that one step
of~\cite{SimonBeig} requires dimension four, and that the family
of asymptotically flat stationary metrics in higher dimensions
might be richer than that in dimension $3+1$,~\cite{MyersPerry}.

\appendix

\section{The infinitesimal invariance group of the constraint equations.}\label{addon}

In this appendix we study the
Lie algebra associated to the group of conformal transformations
preserving the constraint equations~\eq{ceq}. This allows one to
derive identities which shed light on the structure of those
equations.

Suppose that  $\cH=0$ and that the space-time metric $g$ is
rescaled by $\Omega^2$, where, in a Gauss coordinate system
$(t,x^i)$, so that $\hyp=\{t=0\}$, we have
$$\Omega=1+\epsilon \frac{ \psi(x^i)}{j!}t^j\;,$$
for some $j\ge 0$, and $\epsilon>0$ small. In the new Gauss
coordinates $(\bar t, \bar x^i)$ we thus have \beaa
\Omega^2(-dt^2+g_{ij}dx^idx^j )= -d\bar t^2 + \bar g_{k\ell}d\bar
x^kd\bar x^\ell \;,\eeaa  leading to \beal{treq1} &
(\partial_t\bar t)^2 - \bar g_{k\ell}\partial_t \bar x^k\partial_t
\bar x^\ell = \Omega^2\;, & \\
& \partial_t\bar t\partial_i\bar t = \bar g_{k\ell}\partial_i \bar
x^k\partial_t \bar x^\ell \;, & \\ &  g_{ij} = \Omega^{-2} \Big(\bar
g_{k\ell}\partial_i \bar x^k\partial_j  \bar x^\ell - \partial_i
\bar t \partial_j \bar t\Big)\;.&\eeal{treq3} By definition of
Gauss coordinates we have $(\bar t, \bar x^i)=(O(t), x^i+O(t))$,
and also $(\bar t, \bar x^i)=(t+ O(\epsilon), x^i+O(\epsilon))$.
Inserting this in the equations above, matching powers in Taylor
expansions we find
$$\bar t = t+ \epsilon \frac{\psi}{(j+1)!} t^{j+1} + O(\epsilon^2 t^{2j+3})\;,\quad
\bar x^i = x^i +\epsilon\frac{ D^i \psi}{(j+2)!}t^{j+2}
+O(\epsilon^2 t^{2j+2})\;.$$
At the right-hand-side of \eq{treq3} we have variations related to the
fact that all the quantities there are evaluated at the point
$\bar x =x+ \epsilon\times(\cdot)+\ldots$; to first order, this
produces a Lie derivative-type contribution. Next, there are
variations related to the fact that $\bar t =
t+\epsilon\times(\cdot)+\ldots$
 Each
term $  \bar t^i\bar K^{(i)}/i!$ in (twice) the Taylor expansion
of $\bar g_{ij}$ at $t=0$ gives then a contribution to the
right-hand-side of \eq{treq3} equal to $$ \Omega^{-2}(t+ \epsilon
\frac{\psi}{(j+1)!} t^{j+1})^i\frac{\bar
K^{(i)}}{i!}+O(\epsilon^2)= \Big({t^i+
\frac{i-2j-2}{(j+1)!}\epsilon\psi t^{i+j}}\Big)\frac{\bar
K^{(i)}}{i!}+O(\epsilon^2)\;.$$
From this we can calculate the coefficients of an expansion in powers of $t$ of
the right-hand-side of \eq{treq3}; inverting those relations, for
$j=0$ this leads to
\bel{KtransA-1} \bar K^{(i)}= \left\{%
\begin{array}{ll}
    (1-\epsilon(i-2)\psi)K^{(i)} + O(\epsilon^2), & \hbox{$i\ne 2$;} \\
     K^{(i)}- \frac \epsilon 4 \mcL_{D\psi}
g+O(\epsilon^2), & \hbox{$i=2$,} \\
\end{array}%
\right.\ee
 where $\mcL$ denotes a Lie
derivative; note that $\mcL_{D\psi} g$ is twice the Hessian of
$\psi$.

For $j>0$ we obtain instead
\beal{KtransA0}\bar K^{(i)}&=& K^{(i)}\;,\quad 0\le i<j\;,\\\label{KtransA1}\bar K^{(j)}&=& K^{(j)}+
\epsilon \psi g\;,\\\displaystyle\bar K^{(j+i)}&=&
K^{(j+i)}-\frac{(i-2j-2)(j+i)!}{2(j+1)!i!}\epsilon\psi K^{(i)}
\\\nonumber&&+\left\{%
\begin{array}{ll}
     O(\epsilon^2), & \hbox{$1\le i\ne 2$;} \\
    -\frac{\epsilon}{4}
\mcL_{D\psi} g+O(\epsilon^2), & \hbox{$\phantom{1\le\;}i=2$.} \\
\end{array}%
\right.   \eeal{KtransA2} The terms proportional to $\epsilon$ in
those equations describe the desired infinitesimal action.

We can view $\cH_{0\mu}$ as functions of the metric, its
derivatives, and of the symmetric derivatives $D_{(\ell_1\ldots
\ell_s)}K^{(i)}_{k\ell}$, with $i=1,\ldots,n$.
As $\cH_{\mu\nu}$ is a differential operator in the space-time
metric of order $n+1$, the possibly non-trivial contributions
arise from those $D_{(\ell_1\ldots \ell_s)}K^{(i)}_{k\ell}$ for
which $s+i\le n+1$. Differentiating the equations $0=\cH_{0\mu}$
with respect to $\epsilon$, at $\epsilon=t=0$ for $j>0$ one
obtains
\beal{horid} &&\phantom{x} \\\nonumber
0&=&\sum_{s=0}^{n+1-j}\frac{\partial \cH_{0\mu}}{\partial
D_{(\ell_1\ldots \ell_s)}K^{(j)}_{k\ell}} g_{k\ell}
D_{\ell_1\ldots \ell_s}\psi
\\\nonumber&&
-\frac{1}{2}\sum_{i=1}^{n-j}\frac{(i-2j-2)(j+i)!}{(j+1)!i!}\sum_{s=0}^{n+1-i-j}\frac{\partial
\cH_{0\mu}}{\partial D_{(\ell_1\ldots \ell_s)}
K^{(j+i)}_{k\ell}}D_{\ell_1\ldots \ell_s }(\psi
K^{(i)}_{k\ell})\\ &&-\frac{1}{2}\sum_{s=0}^{n-1-j}\frac{\partial
\cH_{0\mu}}{\partial D_{(\ell_1\ldots \ell_s)}
K^{(j+2)}_{k\ell}}D_{\ell_1\ldots \ell_s k \ell}\psi\;.\nonumber\eea
At any point $x$ the derivatives
$D_{\ell_1\ldots \ell_s}\varphi(x)=D_{(\ell_1\ldots
\ell_s)}\varphi(x)$ can be chosen independently, which leads to
various identities. The simplest one is obtained for
$j=n$, then the last two lines give a vanishing
contribution. We parameterise the $K^{(i)}$'s by their trace-free
part and by trace, obtaining
\bel{intid3}0=\frac{\partial \cH_{0\mu}}{\partial D_{\ell_1\ldots
\ell_s} \tr K^{(n)}}\;, \quad s\ge 0\;;\ee this gives a check of
the general form of \eq{ll3}.

When $j=n-1$ the last line in \eq{horid} is zero again, and we obtain
\beaa 0&=&\sum_{s=0}^{2}\frac{\partial \cH_{0\mu}}{\partial
D_{(\ell_1\ldots \ell_s)}K^{(n-1)}_{k\ell}} g_{k\ell}
D_{\ell_1\ldots \ell_s}\psi
\\\nonumber&&
+\frac{1}{2}(2n-1)\sum_{s=0}^{1}\frac{\partial
\cH_{0\mu}}{\partial D_{\ell_1\ldots \ell_s}
K^{(n)}_{k\ell}}D_{\ell_1\ldots \ell_s }(\psi
K^{(1)}_{k\ell})\\&=&\sum_{s=0}^{2}\frac{\partial \cH_{0\mu}}{\partial
D_{\ell_1\ldots \ell_s}\tr K^{(n-1)}}
D_{\ell_1\ldots \ell_s}\psi
\\\nonumber&&
+\frac{1}{2}(2n-1)\Big(\frac{\partial \cH_{0\mu}}{\partial
K^{(n)}_{k\ell}}\psi K^{(1)}_{k\ell}+\frac{\partial
\cH_{0\mu}}{\partial D_{\ell} K^{(n)}_{k\ell}}(
K^{(1)}_{k\ell}D_{\ell}\psi+\psi
D_{\ell}K^{(1)}_{k\ell})\Big)\;.\eeaa The vanishing of the
coefficients in front of the second derivatives of $\psi$ leads to
the identity
\bel{intid4}0=\frac{\partial \cH_{0\mu}}{\partial D_{\ell_1
\ell_2} \tr K^{(n-1)}}\;,\ee consistently with \eq{ll2}.

\bibliographystyle{amsplain}
\bibliography{../references/bartnik,%
../references/myGR,%
../references/newbiblio,%
../references/newbib,%
../references/reffile,%
../references/bibl,%
../references/Energy,%
../references/hip_bib,%
../references/netbiblio}

\end{document}